\documentclass[letterpaper]{raa}
\usepackage{xcolor}
\usepackage{graphicx,times}
\usepackage{natbib}
\usepackage{amssymb,amsmath}
\usepackage{soul}
\bibpunct{(}{)}{;}{a}{}{,}
\usepackage[pagebackref=true]{hyperref}

\begin{document}

\title{Gaussian-Process Emulation of the Redshift-Space Halo Power Spectrum Monopole in Cosmologies with Massive Neutrinos}

\volnopage{{\bf 20XX} Vol.\ {\bf X} No. {\bf XX}, 000--000}
\setcounter{page}{1}

\author{Jixin Gan
   \inst{1,2}
\and Yonghao Feng
   \inst{1,2}
\and Gong-Bo Zhao
   \inst{1,2,3,4}
}

\institute{
National Astronomical Observatories, Chinese Academy of Sciences, Beijing 100101, P. R. China\\
\and
School of Astronomy and Space Science, University of Chinese Academy of Sciences, Beijing 100049, P. R. China\\
\and
Institute for Frontiers in Astronomy and Astrophysics, Beijing Normal University, Beijing 102206, P. R. China\\
\and
Chinese Academy of Sciences South America Center for Astronomy (CASSACA), National Astronomical Observatories, Chinese Academy of Sciences, Beijing 100101, P. R. China\\
\vs \no
{\small Received 20XX Month Day; accepted 20XX Month Day}
}

\abstract{
We present a Gaussian-process (GP) emulator for the monopole of the redshift-space halo power spectrum in $\Lambda$CDM cosmologies with massive neutrinos. The emulator is trained on 1000 COLA simulations distributed in a Latin-hypercube design over the six-dimensional cosmological parameter space $\{\Omega_m h^2,\Omega_b h^2,\Omega_\nu h^2,\sigma_8,h,n_s\}$, with outputs at 11 snapshots spanning $0.5 \le z \le 2.0$. From redshift-space halo catalogues we measure shot-noise-subtracted monopole spectra over $0.01 \le k \le 0.50\,h\,\mathrm{Mpc}^{-1}$. We also generate 1000 fixed-cosmology realizations to estimate the covariance matrix and to construct synthetic data vectors for likelihood tests. On held-out cosmologies, the emulator reproduces the simulated spectra to typically better than $2\%$ across the scales and redshifts considered. Combined with its GP-based estimate of interpolation uncertainty, this speed and accuracy make the emulator well suited to repeated likelihood evaluations in Markov Chain Monte Carlo analyses. The resulting framework provides an efficient route toward neutrino-mass inference from DESI-motivated redshift-space clustering measurements.
}

\authorrunning{J. Gan et al.}
\titlerunning{GP emulator for the redshift-space halo power spectrum}

\maketitle

\section{Introduction}
\label{sect:intro}

Spectroscopic galaxy surveys are among the most powerful probes of late-time cosmology. By mapping the angular positions and redshifts of millions of galaxies over large cosmological volumes, they measure clustering statistics across wide ranges in scale and redshift and thereby constrain both the expansion history and the growth of structure. The statistical power of Stage-IV experiments such as the Dark Energy Spectroscopic Instrument (DESI) makes the galaxy power spectrum a particularly important observable for precision cosmology \citep{DESI2016}. Much of the available information, however, lies on mildly nonlinear scales, where simple analytic models become inadequate. Extracting that information robustly therefore requires forward models that are both accurate and fast enough for repeated use in likelihood-based inference.

A central complication is that spectroscopic clustering is observed in redshift space rather than in real space \citep{Scoccimarro_2004}. Distances along the line of sight are inferred from measured redshifts, which combine the Hubble flow with galaxy peculiar velocities. These peculiar velocities imprint anisotropies on the observed clustering pattern, generating redshift-space distortions (RSD). On large scales, coherent infall enhances the clustering signal through the Kaiser effect \citep{Kaiser1987}, whereas on smaller scales virial motions and nonlinear evolution introduce damping and mode coupling \citep{Hamilton1998}. Accurate modelling of redshift-space clustering must therefore track both the density and velocity fields, as well as their dependence on the underlying cosmological model.

One of the major scientific targets of current large-scale-structure analyses is the absolute neutrino mass scale. Massive neutrinos free-stream over cosmological distances, suppressing the growth of structure below the free-streaming scale and producing characteristic scale- and redshift-dependent signatures in the matter and halo power spectra \citep{Lesgourgues2006,Lesgourgues_2012}. Cosmological data have already placed stringent upper bounds on the sum of neutrino masses \citep{Planck2018VI}, and future spectroscopic surveys should improve those constraints further. Achieving that goal, however, requires accurate modelling of nonlinear clustering in cosmologies with massive neutrinos, where the suppression of power persists into the quasi-nonlinear regime \citep{Bird2012}. In redshift space, neutrino effects enter not only through the density field but also through peculiar velocities and tracer bias, making a robust forward model especially important.

High-fidelity $N$-body simulations remain the most reliable route to nonlinear structure formation, but their computational cost makes them impractical for direct use in Markov Chain Monte Carlo (MCMC) analyses that require many thousands of likelihood evaluations across a multidimensional parameter space. Fast approximate methods therefore play an essential complementary role. In particular, the COmoving Lagrangian Acceleration (COLA) approach provides an effective compromise between speed and accuracy by combining Lagrangian perturbation theory with a particle-mesh treatment of the nonlinear residual evolution \citep{Tassev2013}. Extensions of COLA have made it feasible to generate large mock catalogues in cosmologies with scale-dependent growth, including models with massive neutrinos \citep{Wright2017}; see also \citet{Winther2015} for related fast-simulation methodology. Such methods open the door to simulation suites large enough to train surrogate models.

Simulation-based emulators provide a practical way to bridge the gap between numerical accuracy and inference speed \citep{DeRose_2019, Zhai_2019, Euclid2019, Euclid2021}. The basic idea is to evaluate the target observable on a carefully designed set of training simulations and then interpolate across parameter space with a flexible statistical model. Gaussian-process emulators are especially attractive because they are non-parametric, flexible, and naturally return both a predictive mean and an estimate of interpolation uncertainty \citep{Kwan_2015, Lawrence_2017}. In cosmology, GP-based emulators have already been shown to deliver high-accuracy predictions for a range of clustering observables: for the nonlinear matter power spectrum, see e.g., \citet{Moran_2022, Chen_2025}; for halo clustering, see e.g., \citet{Nishimichi_2019}; and for galaxy clustering, see e.g., \citet{Zhai_2019, Kwan_2023, ruan2026}.
Their probabilistic nature also makes them well suited to likelihood analyses, where interpolation uncertainty can in principle be propagated into the final parameter constraints \citep{Heitmann2009, Rasmussen2006}.

In this work we develop a GP emulator for the monopole of the redshift-space halo power spectrum in $\Lambda$CDM cosmologies with massive neutrinos. Our goal is a forward model that is accurate over a DESI-motivated redshift range and sufficiently fast for repeated likelihood evaluations. To this end, we build a Latin-hypercube training suite of COLA simulations spanning a six-dimensional cosmological parameter space, measure the monopole power spectrum from redshift-space halo catalogues, and train a per-$k$ GP model that predicts $P_0(k)$ as a function of cosmology and redshift. We further generate a dedicated fixed-cosmology suite for covariance estimation and synthetic-data construction, enabling end-to-end tests of an MCMC inference pipeline.

The remainder of this paper is organized as follows. In Section~\ref{sect:method} we describe the simulation suites, the construction of the halo catalogues, the measurement of the redshift-space power spectrum, the GP emulator architecture, and the covariance and MCMC pipelines. In Section~\ref{sect:results} we present emulator validation, the covariance structure, consistency checks between the emulator and the covariance data vector, and the resulting MCMC parameter constraints. In Section~\ref{sect:conclusion} we summarize the present framework, discuss its limitations, and outline natural extensions.

\section{Methodology}
\label{sect:method}

In this section we describe the numerical and statistical ingredients of the emulator. We begin with the simulation suites used for training, validation, and covariance estimation (Section~\ref{subsec:sims}), then present the measurement of the redshift-space halo power spectrum (Section~\ref{subsec:pk}). We next introduce the Gaussian-process emulator architecture (Section~\ref{subsec:emulator}), followed by the covariance construction (Section~\ref{subsec:cov}) and the MCMC inference pipeline (Section~\ref{subsec:mcmc}).

\subsection{Simulation Suites}
\label{subsec:sims}

\subsubsection{COLA Simulations}

We generate two ensembles of approximate $N$-body simulations using the \textsc{COLA} (COmoving Lagrangian Acceleration) method \citep{Tassev2013,Winther2015,Wright2017}. The first is a Latin-hypercube-sampling (LHS) suite of 1000 simulations used to train and validate the emulator. The second is a fixed-cosmology suite of 1000 independent realizations used to estimate the covariance matrix adopted in the likelihood analysis. Both suites share the same numerical setup, so differences between them arise only from cosmological sampling and random initial conditions.

All simulations are evolved in a periodic cubic box of side length $L_\mathrm{box}=1024\,h^{-1}\mathrm{Mpc}$ with $1024^3$ cold-dark-matter particles, from the initial redshift $z_\mathrm{init}=19$ to $z=0.5$. Dark matter haloes are identified with the Friends-of-Friends algorithm \citep{1982ApJ...257..423H,1983ApJS...52...61G} as implemented in the \textsc{Matchmaker} halo finder \citep{matchmaker}, using a linking length of $b=0.2$. Halo masses are defined as the sum of the member-particle masses.

\subsubsection{LHS Training Suite}
\label{subsubsec:lhs}

The LHS suite is designed to sample efficiently the parameter space of flat $\Lambda$CDM cosmologies with massive neutrinos \citep{LHS}. We vary six parameters: the total matter density $\Omega_m h^2$, the baryon density $\Omega_b h^2$, the neutrino density $\Omega_\nu h^2$ (related to the summed neutrino mass through $\sum m_\nu \approx 93.14\,\Omega_\nu h^2\,\mathrm{eV}$), the amplitude of matter fluctuations $\sigma_8$, the dimensionless Hubble parameter $h$, and the scalar spectral index $n_s$. The prior ranges adopted for these parameters are listed in Table~\ref{tab:params}; they define both the Latin-hypercube design and the priors used in the MCMC analysis.

\begin{figure}[htbp]
  \begin{minipage}[t]{0.48\linewidth}
    \centering
    \includegraphics[width=\linewidth]{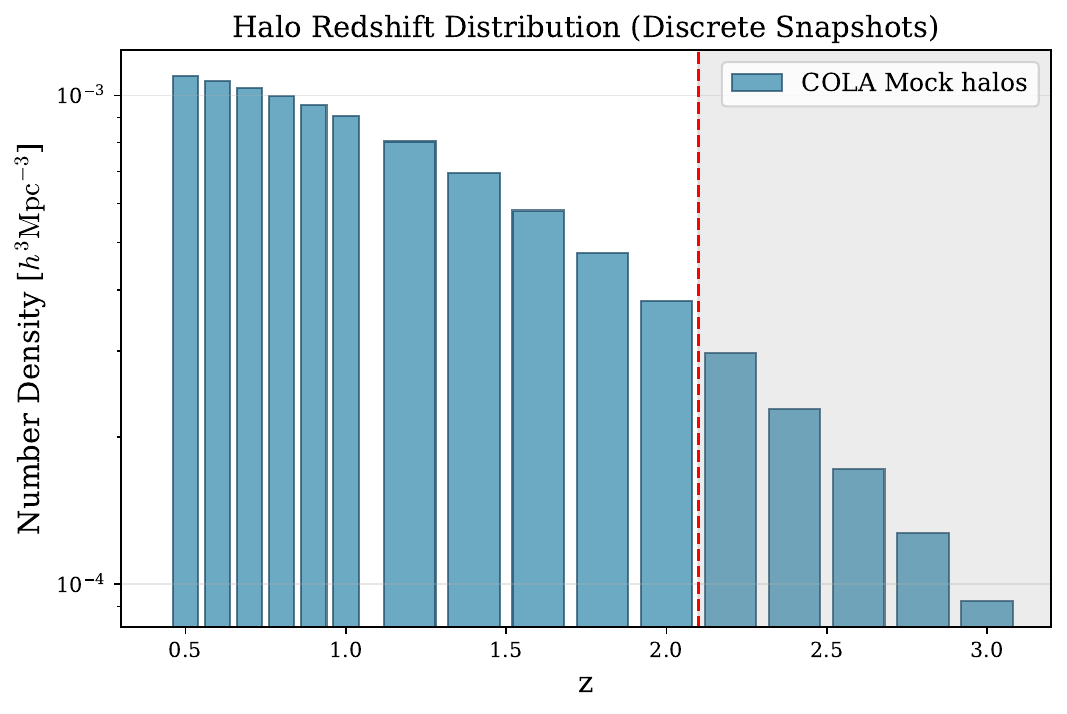}
    \centerline{(a) LHS suite}
  \end{minipage}%
  \hfill
  \begin{minipage}[t]{0.48\linewidth}
    \centering
    \includegraphics[width=\linewidth]{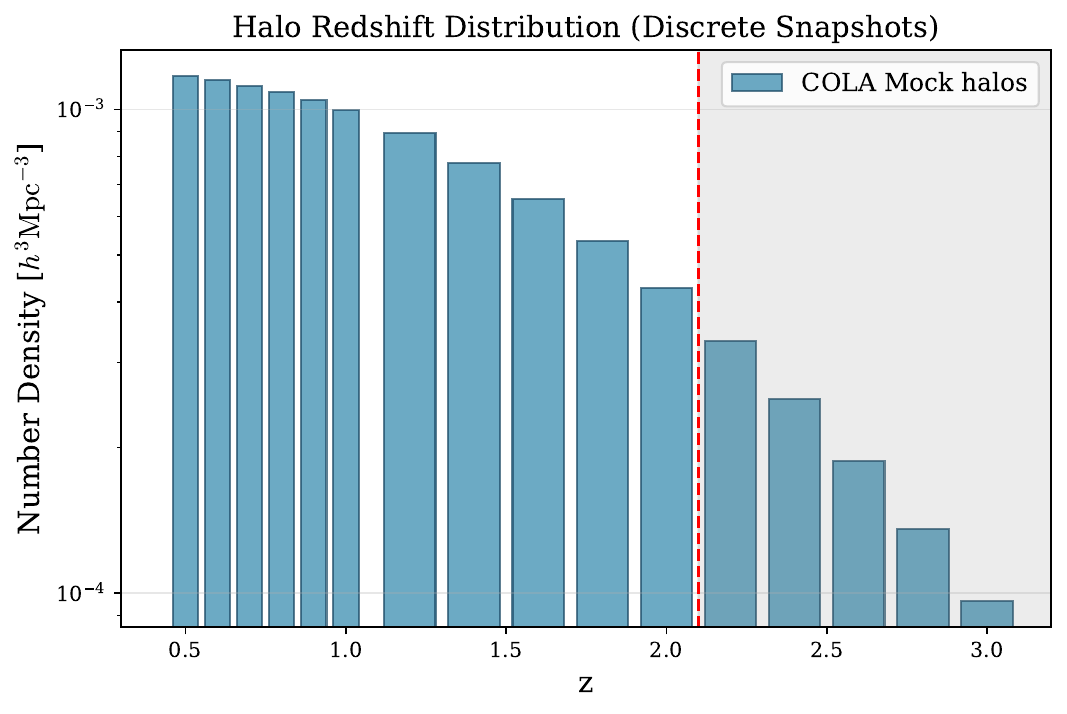}
    \centerline{(b) Covariance suite}
  \end{minipage}
  \caption{Halo number density as a function of redshift for (a) the LHS training suite (median across 1000 cosmologies) and (b) the covariance suite at the fiducial cosmology. The present analysis is restricted to $z\leq 2.0$; the red dashed line marks this cut and the gray shaded region indicates the excluded range $z>2.0$. Beyond the cut, the number density drops below $\sim 10^{-4}\,h^3\mathrm{Mpc}^{-3}$ and shot noise dominates the measured power spectrum.}
  \label{fig:ndens}
\end{figure}

We draw 1000 design points in this six-dimensional space using a maximin Latin-hypercube design generated with the \texttt{pyDOE2} library \citep{pyDOE2}. Each cosmology is evolved to 11 snapshots at redshifts $z=0.5$, $0.6$, $0.7$, $0.8$, $0.9$, $1.0$, $1.2$, $1.4$, $1.6$, $1.8$, and $2.0$. Although outputs are available up to $z=3.0$, we restrict the present analysis to $z\leq 2.0$. Beyond that limit the halo number density falls below $\sim 10^{-4}\,h^3\mathrm{Mpc}^{-3}$ (Fig.~\ref{fig:ndens}), corresponding to fewer than $\sim 10^4$ haloes per realization, so shot noise increasingly dominates the measured power spectrum.

All 1000 LHS simulations share a single random seed for the initial conditions, so differences among them arise purely from the variation in cosmological parameters. This choice suppresses realization-to-realization scatter in the training targets and helps the emulator isolate the cosmology dependence of the power spectrum. A consequence is that the emulator prediction at any given cosmology implicitly retains the particular large-scale-mode configuration of the shared seed; we return to this point in Section~\ref{subsec:consistency}.

The adopted redshift range is motivated by the DESI luminous red galaxy sample \citep{Zhou_2023}, whose primary target selection lies roughly in the interval $0.4\lesssim z\lesssim 1.1$, while still allowing straightforward extension to higher-redshift samples such as emission-line galaxies and quasars. Figure~\ref{fig:ndens} shows the redshift evolution of the halo number density in the LHS and covariance suites.

\subsubsection{Covariance Suite}
\label{subsubsec:covsims}

The covariance suite consists of 1000 independent realizations at the fiducial cosmology listed in Table~\ref{tab:params}, each generated from a different random seed. These simulations use the same box size, mass resolution, redshift outputs, and halo-finding procedure as the LHS suite. They are therefore well suited to estimate the covariance matrix of the power spectrum at each redshift. We use the power-spectrum mean averaged over all 1000 realizations as the synthetic data vector in the MCMC analysis described in Section~\ref{subsec:mcmc}.

\begin{table}[htbp]
    \centering
    \caption{Cosmological parameter ranges of the LHS training suite. The ``Range'' column lists the uniform bounds adopted for both the Latin-hypercube design and the MCMC priors. The ``Fiducial'' column gives the cosmology used for the covariance suite and for the synthetic data vector in the likelihood analysis.}
    \label{tab:params}
    \begin{tabular}{lccc}
        \hline\hline
        Parameter & Symbol & Range & Fiducial \\
        \hline
        Total matter density  & $\Omega_m h^2$   & $[0.120,\,0.155]$   & $0.1400$ \\
        Baryon density        & $\Omega_b h^2$   & $[0.0215,\,0.0235]$ & $0.02206$ \\
        Neutrino density      & $\Omega_\nu h^2$ & $[0.00017,\,0.01]$  & $0.0007$ \\
        Fluctuation amplitude & $\sigma_8$       & $[0.7,\,0.9]$       & $0.8110$ \\
        Hubble parameter      & $h$               & $[0.55,\,0.85]$     & $0.6710$ \\
        Scalar spectral index & $n_s$             & $[0.85,\,1.05]$     & $0.9665$ \\
        \hline
    \end{tabular}
\end{table}

\subsection{Power Spectrum Measurement}
\label{subsec:pk}

The observable emulated in this work is the monopole of the redshift-space halo power spectrum, $P_0(k)$. Spectroscopic surveys infer radial distances from observed redshifts, so peculiar velocities distort the apparent clustering pattern along the line of sight and encode additional information about the growth of structure \citep{Kaiser1987,Hamilton1998}. Since the emulator is ultimately intended for DESI-motivated analyses, we work directly in redshift space throughout.

\subsubsection{Redshift-Space Distortions}
\label{subsubsec:rsd}

To construct redshift-space halo catalogues, we displace each halo along a fixed line-of-sight direction, chosen here to be the $z$-axis of the simulation box, according to
\begin{equation}
\label{eq:rsd}
    \mathbf{s} = \mathbf{r} + \frac{v_z}{aH(a)}\,\hat{\mathbf{z}}\,,
\end{equation}
where $\mathbf{r}$ is the real-space position, $v_z$ is the line-of-sight component of the halo peculiar velocity, $a$ is the scale factor, and $H(a)$ is the Hubble parameter at the redshift of the snapshot. This plane-parallel (distant-observer) approximation is standard for periodic simulation boxes and is adequate for the scales considered here.

\subsubsection{Estimator and Binning}
\label{subsubsec:pestimator}

We measure the power spectrum with the \textsc{Nbodykit} package \citep{Hand2018} using a fast-Fourier-transform-based estimator. Haloes are assigned to a regular mesh of $N_\mathrm{mesh}^3=512^3$ cells using a third-order interpolation scheme. The resulting density contrast field, $\delta(\mathbf{x})=n(\mathbf{x})/\bar{n}-1$, is Fourier transformed and the three-dimensional power spectrum is decomposed into Legendre multipoles via
\begin{equation}
\label{eq:multipole}
    P_\ell(k)=\frac{2\ell+1}{N_k}\sum_{|\mathbf{k}|\in k\text{-bin}} \mathcal{L}_\ell(\hat{\mathbf{k}}\cdot\hat{\mathbf{z}})\,|\delta(\mathbf{k})|^2\,,
\end{equation}
where $\mathcal{L}_\ell$ is the Legendre polynomial of order $\ell$, $\hat{\mathbf{z}}$ is the line-of-sight direction, and $N_k$ is the number of Fourier modes in each bin. In this work we restrict attention to the monopole ($\ell=0$), leaving higher multipoles such as the quadrupole and hexadecapole to future work.

We use linear $k$ bins of width $\Delta k=0.005\,h\,\mathrm{Mpc}^{-1}$. The resulting data vector contains 97 bin centres spanning $k\simeq 0.013$ to $0.493\,h\,\mathrm{Mpc}^{-1}$, corresponding to the nominal interval $0.01\leq k\leq 0.50\,h\,\mathrm{Mpc}^{-1}$. The lower end lies safely above the fundamental mode of the box, $2\pi/L_\mathrm{box}\approx 0.006\,h\,\mathrm{Mpc}^{-1}$, while the upper end reaches into the mildly nonlinear regime and balances the gain in information against the declining accuracy of approximate dynamics at large $k$.

\subsubsection{Shot Noise Correction}
\label{subsubsec:shotnoise}

The raw power spectrum includes a Poisson shot-noise contribution arising from the discrete sampling of the density field by a finite number of haloes \citep{Cohn_2006}. Under the Poisson approximation, this contribution is scale independent and equal to the inverse mean halo number density,
\begin{equation}
\label{eq:shotnoise}
    P_\mathrm{shot}=\frac{1}{\bar{n}}=\frac{V_\mathrm{box}}{N_\mathrm{halo}}\,,
\end{equation}
where $V_\mathrm{box}=L_\mathrm{box}^3$ is the simulation volume and $N_\mathrm{halo}$ is the total number of haloes in the catalogue. We subtract this contribution from the measured monopole,
\begin{equation}
\label{eq:pkcorr}
    P_0(k) \leftarrow P_{0,\mathrm{raw}}(k)-P_\mathrm{shot}\,,
\end{equation}
and use the shot-noise-subtracted spectrum as the emulator target. For halo samples, the true stochastic contribution need not be perfectly Poissonian, especially for massive tracers that can exhibit sub-Poisson behaviour \citep{Seljak2009,Baldauf2013}. For the present work, however, simple Poisson subtraction provides a consistent and practical baseline, and any residual deviations are absorbed into emulator training and validation.

The resulting set of $1000\times 11=11{,}000$ monopole power spectra (1000 cosmologies at 11 redshifts) forms the full dataset used for emulator training and validation.

\subsection{Gaussian-Process Emulator}
\label{subsec:emulator}

Directly evaluating the halo power spectrum at arbitrary cosmological parameters would require either new simulations or repeated runs of an approximate solver, which is still too expensive for iterative Bayesian inference. We therefore construct a fast surrogate model based on Gaussian-process (GP) regression \citep{Rasmussen2006}. GP emulators are well suited to this task because they interpolate smoothly across parameter space and provide both a predictive mean and a calibrated uncertainty estimate. Such methods have been used successfully in cosmology for nonlinear-structure observables, including power-spectrum emulation \citep{Heitmann2009}.

\subsubsection{Gaussian-Process Regression}
\label{subsubsec:gp_formalism}

A Gaussian process defines a distribution over functions, fully specified by a mean function $m(\mathbf{x})$ and a covariance (kernel) function $\kappa(\mathbf{x},\mathbf{x}')$ \citep{Rasmussen2006}. Given a training set $\mathcal{D}=\{(\mathbf{x}_i,y_i)\}_{i=1}^{N_\mathrm{train}}$, where $\mathbf{x}_i$ denotes the input cosmological parameters and $y_i$ the corresponding logarithmic power-spectrum value at a fixed wavenumber and redshift, the GP posterior predictive distribution at a new test point $\mathbf{x}_*$ is Gaussian with mean and variance
\begin{align}
\label{eq:gp_mean}
    \mu(\mathbf{x}_*) &= m(\mathbf{x}_*) + \mathbf{k}_*^\top\left[\mathbf{K}+\sigma_n^2\mathbf{I}\right]^{-1}(\mathbf{y}-\mathbf{m})\,, \\
\label{eq:gp_var}
    \sigma^2(\mathbf{x}_*) &= \kappa(\mathbf{x}_*,\mathbf{x}_*) - \mathbf{k}_*^\top\left[\mathbf{K}+\sigma_n^2\mathbf{I}\right]^{-1}\mathbf{k}_*\,,
\end{align}
where $\mathbf{K}$ is the $N_\mathrm{train}\times N_\mathrm{train}$ training covariance matrix with entries $K_{ij}=\kappa(\mathbf{x}_i,\mathbf{x}_j)$, $\mathbf{k}_*=[\kappa(\mathbf{x}_*,\mathbf{x}_1),\ldots,\kappa(\mathbf{x}_*,\mathbf{x}_{N_\mathrm{train}})]^\top$ is the cross-covariance vector between the test point and the training set, $\mathbf{m}$ is the vector of mean-function values evaluated at the training inputs, and $\sigma_n^2$ is a noise variance term that accounts for stochasticity in the training targets (Section~\ref{subsubsec:noise}). When transformed back from log space, the emulator prediction is $P_0^{\mathrm{emu}}(k;\mathbf{x}_*)=\exp[\mu(\mathbf{x}_*)]$.

\subsubsection{Kernel and Mean Function}
\label{subsubsec:kernel}

We adopt a constant mean function,
\begin{equation}
\label{eq:mean}
    m(\mathbf{x})=c\,,
\end{equation}
where $c$ is a free hyperparameter learned during training. For the covariance function we use a radial-basis-function (RBF) kernel, also known as the squared-exponential kernel, with automatic relevance determination (ARD) \citep{Rasmussen2006}:
\begin{equation}
\label{eq:kernel}
    \kappa(\mathbf{x},\mathbf{x}')=\sigma_f^2\exp\!\left[-\frac{1}{2}\sum_{d=1}^{D}\frac{(x_d-x_d')^2}{\ell_d^2}\right],
\end{equation}
where $\sigma_f^2$ is the output-scale variance, $\ell_d$ is the characteristic length scale associated with the $d$th input dimension, and $D=6$ is the number of cosmological parameters. The ARD parametrization assigns an independent length scale to each parameter, allowing the emulator to capture anisotropic sensitivity across parameter space. Shorter length scales indicate directions along which the power spectrum varies more rapidly, while longer length scales correspond to weaker parameter dependence. The complete hyperparameter set is therefore $\boldsymbol{\theta}=\{c,\sigma_f^2,\ell_1,\ldots,\ell_6,\sigma_n^2\}$, comprising $D+3=9$ free parameters.

\subsubsection{Training Procedure}
\label{subsubsec:training}

The GP hyperparameters are determined by maximizing the log marginal likelihood (LML) of the training data,
\begin{equation}
\label{eq:lml}
    \log p(\mathbf{y}\mid\mathbf{X},\boldsymbol{\theta}) = -\frac{1}{2}\mathbf{y}^\top\mathbf{K}_\theta^{-1}\mathbf{y} - \frac{1}{2}\log|\mathbf{K}_\theta| - \frac{N_\mathrm{train}}{2}\log(2\pi)\,,
\end{equation}
where $\mathbf{K}_\theta=\mathbf{K}+\sigma_n^2\mathbf{I}$. For notational compactness, the contribution of the mean function is absorbed into $\mathbf{y}$. Maximizing the LML naturally balances goodness of fit against model complexity and therefore provides an intrinsic regularization against overfitting.

We implement the emulator with the \textsc{GPyTorch} library \citep{Gardner2018}, which provides efficient GPU-accelerated kernel operations and linear algebra routines for GP inference. Hyperparameter optimization is performed with the Adam optimizer \citep{Kingma2014} for a fixed number of iterations until the marginal likelihood converges across the full set of trained models.

\subsubsection{Per-Bin Emulation Strategy}
\label{subsubsec:perbin}

We train one independent GP for each $k$ bin at each of the 11 redshift snapshots. This per-bin strategy has several practical advantages. First, each GP is trained on a modest dataset of $N_\mathrm{train}=800$ points in a six-dimensional input space, for which exact GP inference remains computationally tractable. Second, the kernel hyperparameters are free to adapt to the scale-dependent response of the power spectrum to cosmological parameters. Third, the models can be trained independently and therefore in parallel, which keeps the total wall-clock cost manageable on modern GPU hardware.

The main trade-off is that correlations between neighbouring $k$ bins are not modelled explicitly within the emulator itself. In practice, however, the underlying power spectrum is a smooth function of scale, and neighbouring GPs are trained on closely related targets. As a result, the reconstructed spectra remain smooth and physically well behaved across the full $k$ range considered here.

We note that because a separate GP is trained at each redshift of the snapshots, the current emulator does not interpolate among redshifts. Predictions are therefore limited to the 11 discrete output times of the simulation suite. Extending the emulator to accept redshifts as a continuous input parameter is a natural next step and is left for future work.

\subsubsection{Input and Output Preprocessing}
\label{subsubsec:preprocess}

Before training, we standardize each input parameter to zero mean and unit variance over the training set. This improves numerical conditioning and places all dimensions on comparable scales for kernel optimization. For the output, we emulate the natural logarithm of the power spectrum, $y_i=\ln P_0(k;\mathbf{x}_i)$, rather than $P_0$ itself. Working in log space guarantees positive predictions after exponentiation, compresses the dynamic range of the targets, and typically improves the stationarity of the GP model across the full $k$ range.

\subsubsection{Noise Treatment}
\label{subsubsec:noise}

As discussed in Section~\ref{subsubsec:lhs}, each training spectrum is measured from a single simulation realization and therefore contains stochastic fluctuations from sample variance and the finite number of haloes. The noise hyperparameter $\sigma_n^2$ in Eq.~(\ref{eq:gp_mean}) absorbs this effective numerical noise. During training it is optimized jointly with the kernel hyperparameters through the marginal likelihood, so the GP learns the appropriate degree of smoothing directly from the data. A nonzero $\sigma_n^2$ means that the emulator does not interpolate the training points exactly, which is desirable when the targets themselves are noisy. The predictive variance in Eq.~(\ref{eq:gp_var}) correspondingly provides a conservative uncertainty estimate for the emulator output.

\subsubsection{Training and Validation Split}
\label{subsubsec:split}

\begin{figure*}[htbp]
  \centering
  \includegraphics[width=\textwidth]{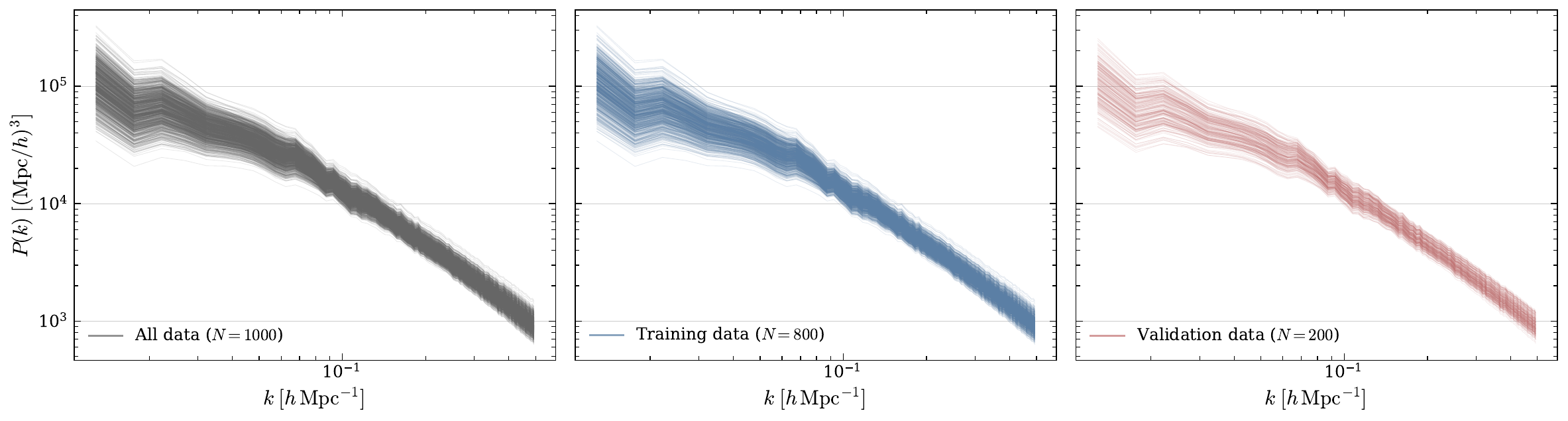}
  \caption{Redshift-space halo power-spectrum monopole $P_0(k)$ at $z=0.5$ for the full halo sample. \textit{Left:} all 1000 LHS simulations (gray). \textit{Middle:} the 800 simulations used for GP training (blue). \textit{Right:} the 200 simulations held out for validation (red). The spread in amplitude at each wavenumber reflects the range of cosmological parameters sampled by the Latin-hypercube design (Table~\ref{tab:params}). The validation set spans a range comparable to that of the training set, confirming that the random 80--20 split does not introduce a selection bias.}
  \label{fig:training_data}
\end{figure*}

Out of the 1000 LHS simulations, we randomly select 800 ($80\%$) for training and reserve the remaining 200 ($20\%$) for validation. The split is performed once and held fixed across all $k$ bins and redshifts, ensuring that every GP is tested on the same held-out cosmologies. We quantify predictive performance using standard diagnostics such as the coefficient of determination $R^2$ and the fractional residual $(P_\mathrm{pred}-P_\mathrm{true})/P_\mathrm{true}$. Once trained, the emulator can predict the full monopole spectrum for a given cosmology and redshift at negligible cost compared with running a new simulation, making it suitable for repeated likelihood evaluations. The resulting training and validation spectra at $z=0.5$ are shown in Fig.~\ref{fig:training_data}.

\subsection{Covariance Matrix Estimation}
\label{subsec:cov}

Likelihood-based parameter inference requires a covariance matrix that captures the uncertainties and inter-bin correlations of the measured monopole spectrum. We estimate this covariance empirically from the 1000 fixed-cosmology realizations described in Section~\ref{subsubsec:covsims}. At each redshift, we measure the shot-noise-subtracted monopole spectrum in every realization, compute the ensemble mean,
\begin{equation}
\label{eq:mean_pk}
    \bar{P}_i = \frac{1}{N_\mathrm{r}}\sum_{n=1}^{N_\mathrm{r}} P_i^{(n)}\,,
\end{equation}
and form the sample covariance,
\begin{equation}
\label{eq:cov}
    C_{ij}=\frac{1}{N_\mathrm{r}-1}\sum_{n=1}^{N_\mathrm{r}}\left[P_i^{(n)}-\bar{P}_i\right]\left[P_j^{(n)}-\bar{P}_j\right],
\end{equation}
where $N_\mathrm{r}=1000$ and the indices $i$ and $j$ run over the 97 $k$ bins at a fixed redshift.

The ensemble-averaged spectrum serves two purposes. First, it provides the synthetic data vector used in the mock MCMC tests. Second, together with the sample covariance, it defines a self-consistent likelihood setup matched to the same simulation specifications, halo selection, and power-spectrum estimator used to train the emulator. Because the covariance suite contains far more realizations than the dimensionality of the data vector, the resulting covariance estimate is sufficiently well sampled for the present proof-of-principle study.

\subsection{MCMC}
\label{subsec:mcmc}

We use the emulator as the theory engine in a standard Bayesian parameter-inference pipeline. At each step of the Markov chain, a trial cosmological parameter vector $\{\Omega_m h^2,\Omega_b h^2,\Omega_\nu h^2,\sigma_8,h,n_s\}$ is passed to the emulator, which returns the model prediction for the redshift-space monopole at the chosen redshift. In the tests presented below, each redshift snapshot is analysed independently. We compare the emulator prediction with the synthetic data vector derived from the covariance suite through a Gaussian likelihood,
\begin{equation}
\label{eq:like}
    -2\ln \mathcal{L}(\boldsymbol{\theta}) = \left[\mathbf{d}-\mathbf{m}(\boldsymbol{\theta})\right]^\top \mathbf{C}^{-1} \left[\mathbf{d}-\mathbf{m}(\boldsymbol{\theta})\right],
\end{equation}
where $\mathbf{d}$ is the mean monopole from the covariance suite, $\mathbf{m}(\boldsymbol{\theta})$ is the emulated prediction, and $\mathbf{C}$ is the sample covariance matrix.

The priors are taken to be uniform over the same parameter ranges used to generate the LHS design (Table~\ref{tab:params}), ensuring that posterior exploration remains inside the domain covered by the emulator training set. In the present implementation, the exponentiated GP predictive mean is used as the theory vector in the likelihood, while the GP predictive variance is monitored separately as a diagnostic of interpolation quality rather than folded into $\mathbf{C}$. This setup allows us to test, in a controlled environment, whether the emulator can recover the fiducial cosmology from synthetic data without introducing significant bias.

Replacing an explicit simulation call with an emulator evaluation drastically reduces the cost of each likelihood computation. As a result, the total runtime of the inference is dominated by operations on the data vector and covariance matrix rather than by forward modelling. This speed gain is essential for practical MCMC applications in multidimensional cosmological parameter spaces.

\section{Results}
\label{sect:results}

In this section we present the main results of the emulator framework. We first assess the predictive accuracy of the trained GP models on validation cosmologies (Section~\ref{subsec:emu_results}). We then examine the covariance matrix estimated from the fixed-cosmology suite (Section~\ref{subsec:cov_results}), followed by a consistency check between the emulator prediction and the covariance data vector (Section~\ref{subsec:consistency}). Finally, we present MCMC parameter constraints obtained using the emulator as the theory model (Section~\ref{subsec:mcmc_results}).

\subsection{Emulator Validation}
\label{subsec:emu_results}

We validate the emulator by comparing its predictions against the 200 LHS simulations that were not used during training. Because the validation cosmologies are drawn from the same Latin hypercube as the training set and share the same initial random seed, this test isolates GP interpolation accuracy across parameter space rather than the additional realization-to-realization scatter associated with new phases.

To illustrate the emulator performance across the neutrino-mass range, we select two representative validation cosmologies near opposite ends of the prior: one with the highest neutrino density in the validation set, $\Omega_\nu h^2=0.0099$, and one with the lowest value, $\Omega_\nu h^2=0.0002$. These cases bracket the sampled range and test whether the GP captures the scale-dependent suppression of power associated with massive neutrinos.

\begin{figure*}[htbp]
  \centering
  \begin{minipage}[t]{\textwidth}
    \centering
    \includegraphics[width=\textwidth]{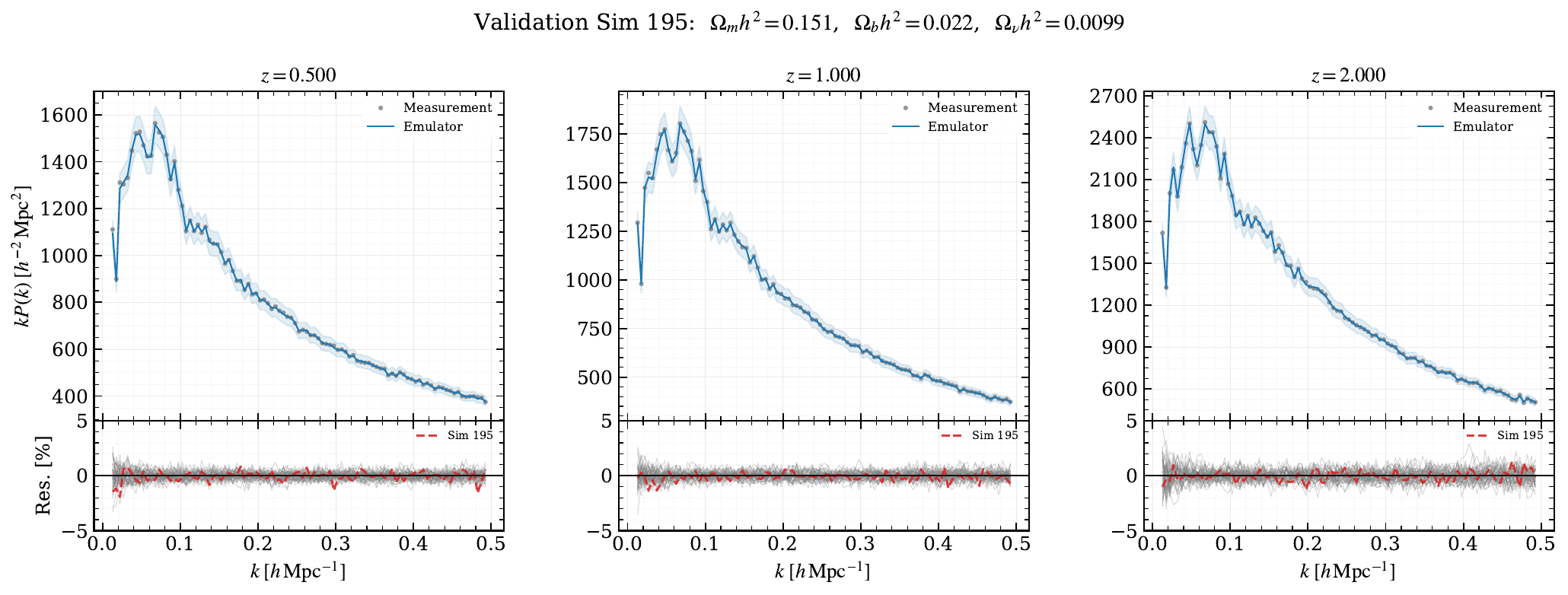}
    \centerline{(a) High neutrino mass}
  \end{minipage}
  \vspace{6pt}
  \begin{minipage}[t]{\textwidth}
    \centering
    \includegraphics[width=\textwidth]{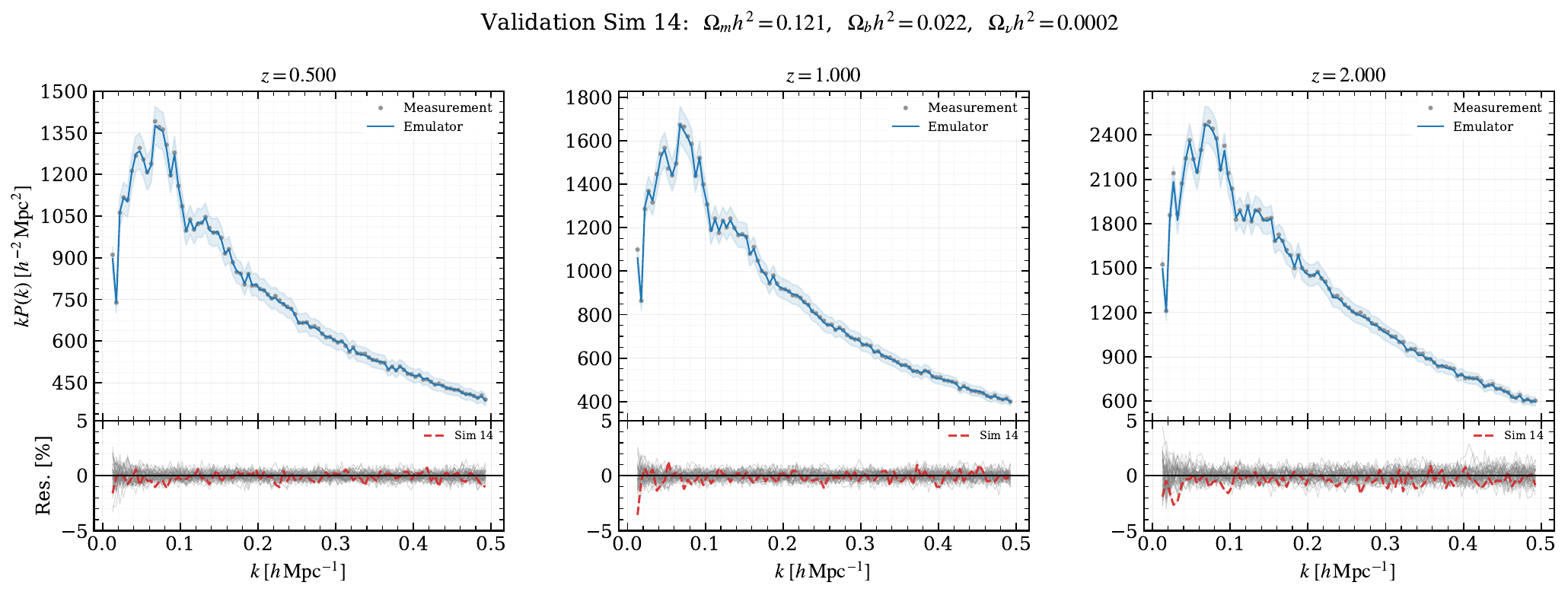}
    \centerline{(b) Low neutrino mass}
  \end{minipage}
  \caption{Emulator prediction (blue) versus simulation truth (gray points) for two representative validation cosmologies at $z=0.5$, $1.0$, and $2.0$ (redshift-space monopole, full halo sample). (a) High-neutrino-mass case with $\Omega_m h^2=0.151$, $\Omega_b h^2=0.022$, and $\Omega_\nu h^2=0.0099$. (b) Low-neutrino-mass case with $\Omega_m h^2=0.121$, $\Omega_b h^2=0.022$, and $\Omega_\nu h^2=0.0002$. Upper sub-panels show $kP_0(k)$: the blue line is the emulator prediction with $\pm 1\sigma$ GP uncertainty (shaded band), and the gray points are the simulation measurement. Lower sub-panels show the fractional residual $(P_\mathrm{pred}-P_\mathrm{true})/P_\mathrm{true}$ for 50 randomly selected validation samples. Residuals remain within $\sim 5\%$ across all scales and redshifts shown.}
  \label{fig:emu_pred}
\end{figure*}

Figures~\ref{fig:emu_pred}(a) and \ref{fig:emu_pred}(b) compare the emulator prediction and simulation truth for the high- and low-neutrino-mass examples at three redshifts, $z=0.5$, $1.0$, and $2.0$. In each case, the upper sub-panel shows the predicted $kP_0(k)$ (blue) together with the simulation measurement (gray points), while the shaded band indicates the GP $\pm 1\sigma$ predictive uncertainty. The lower sub-panel displays the fractional residual $(P_\mathrm{pred}-P_\mathrm{true})/P_\mathrm{true}$ for 50 randomly selected validation cosmologies (gray), with the highlighted example shown in red. Across all scales and redshifts, the residuals are typically within $\sim 2\%$ and remain well inside the $\pm 5\%$ envelope, with somewhat larger scatter at the lowest wavenumbers where the number of independent Fourier modes is smallest.

These comparisons show that the emulator interpolates reliably across the full six-dimensional parameter space, maintaining $\lesssim 2\%$ accuracy from the lowest to the highest neutrino masses, from $z=0.5$ to $z=2.0$, and from the largest scales down to $k\approx 0.5\,h\,\mathrm{Mpc}^{-1}$. The sub-percent accuracy reached at intermediate and high $k$ is particularly encouraging. In this regime the power spectrum is sensitive to the interplay among nonlinear growth, neutrino free streaming, and redshift-space distortions, so the response surface is most complex. The fact that the GP residuals remain small and unbiased across all six redshifts indicates that the per-$k$-bin emulation strategy, combined with the ARD kernel, captures the relevant parameter dependence without overfitting the noise in the training spectra.

\subsection{Covariance Matrix}
\label{subsec:cov_results}

The covariance matrix used in the likelihood analysis is estimated from 1000 independent realizations at the fiducial cosmology (Section~\ref{subsubsec:covsims}). We examine its structure at six redshifts spanning the analysis range: $z=0.5$, $0.7$, $0.9$, $1.2$, $1.6$, and $2.0$.

\begin{figure*}[htbp]
  \centering
  \includegraphics[width=\textwidth]{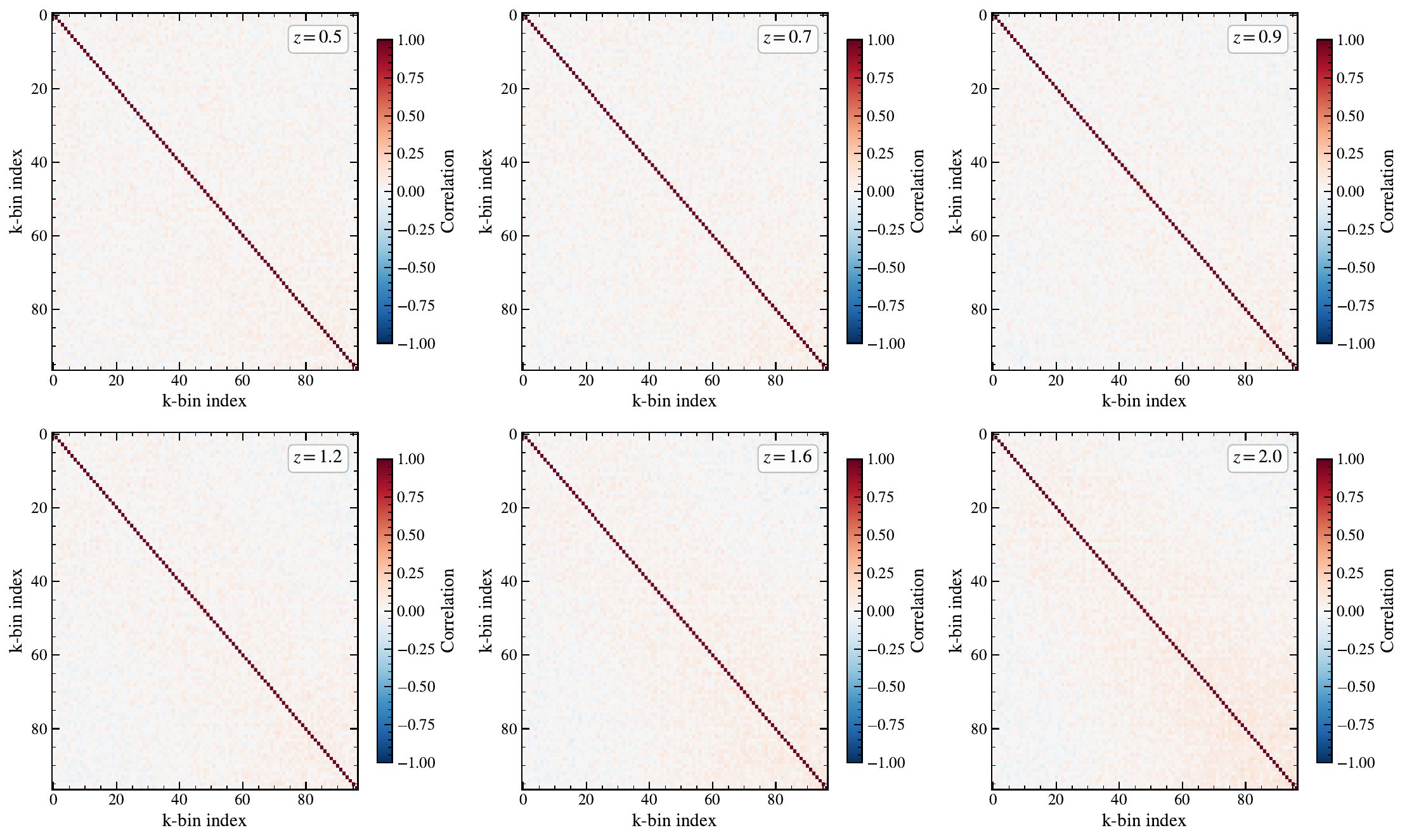}
  \caption{Correlation matrices of the redshift-space monopole power spectrum at six redshifts ($z=0.5$, $0.7$, $0.9$, $1.2$, $1.6$, $2.0$), estimated from 1000 fixed-cosmology realizations (full halo sample). All matrices are close to diagonal at the adopted $k$-bin spacing, with only mild off-diagonal structure at the lowest wavenumbers. The correlation pattern remains qualitatively stable across the redshift range.}
  \label{fig:corr_matrices}
\end{figure*}

Figure~\ref{fig:corr_matrices} presents the corresponding correlation coefficient matrices, $r_{ij}=C_{ij}/\sqrt{C_{ii}C_{jj}}$, at these six redshifts. All matrices are close to diagonal, with only mild off-diagonal structure, indicating that inter-bin correlations are weak at the adopted bin spacing of $\Delta k=0.005\,h\,\mathrm{Mpc}^{-1}$.

\begin{figure*}[htbp]
  \centering
  \includegraphics[width=\textwidth]{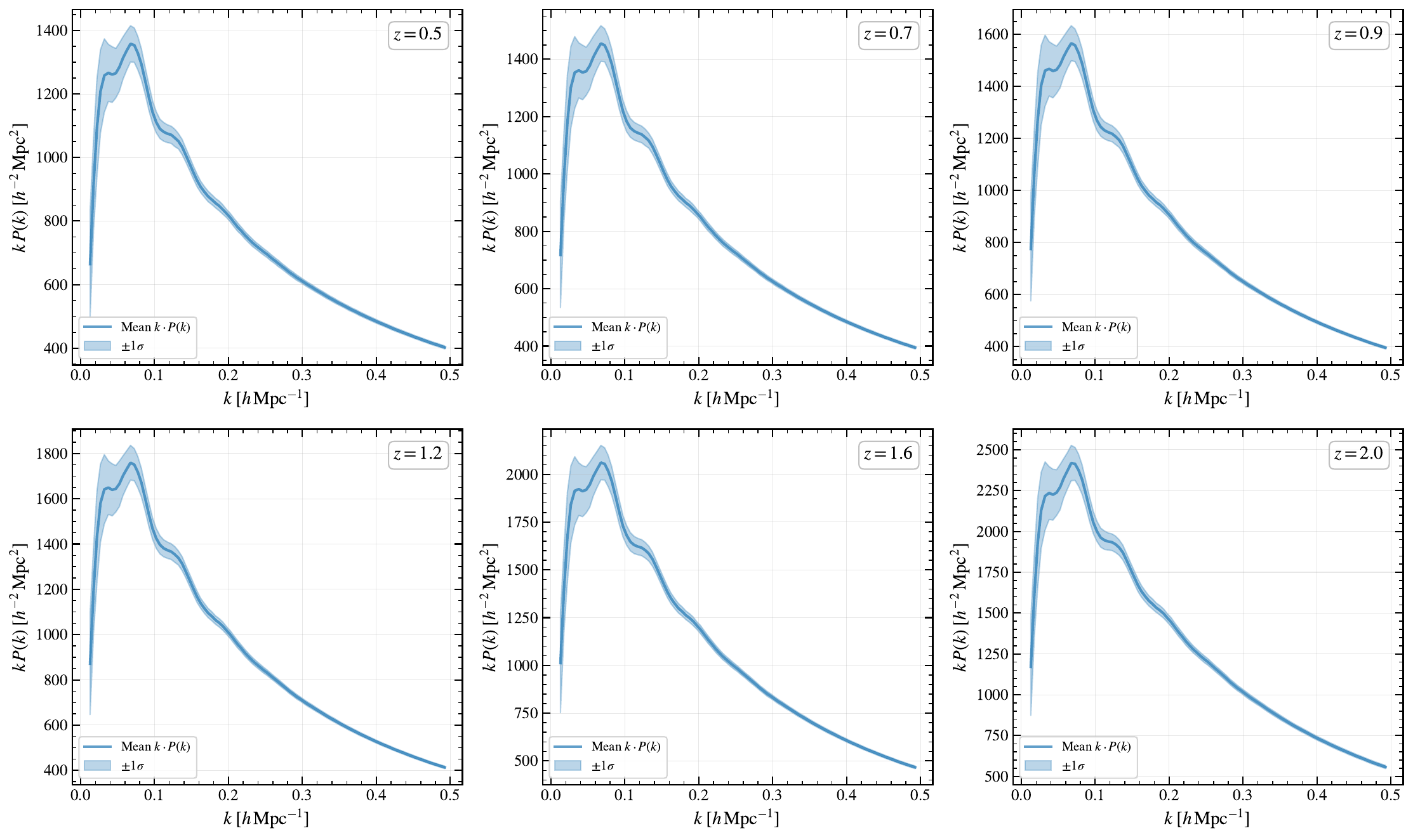}
  \caption{Mean monopole power spectrum $kP_0(k)$ (solid line) and $\pm 1\sigma$ scatter (shaded band) from 1000 fixed-cosmology realizations at six redshifts, corresponding to the correlation matrices in Fig.~\ref{fig:corr_matrices}. The relative scatter is largest at low $k$, where cosmic variance dominates. The BAO features are clearly resolved at all redshifts.}
  \label{fig:mean_kpk}
\end{figure*}

Figure~\ref{fig:mean_kpk} shows the corresponding mean monopole spectrum, plotted as $kP_0(k)$, together with the $\pm 1\sigma$ scatter from the 1000 realizations. The relative scatter is largest at low $k$, where cosmic variance dominates, and decreases steadily toward higher wavenumbers as the $1\sigma$ band narrows with increasing $k$. This ensemble-averaged spectrum at each redshift serves as the synthetic data vector in the MCMC analysis.

\subsection{Emulator--Data Vector Consistency}
\label{subsec:consistency}

Before parameter inference, we examine the agreement between the emulator prediction at the fiducial cosmology and the mean data vector from the covariance suite. The 1000 LHS training simulations share a single random seed, whereas the 1000 covariance realizations each use an independent seed. As a result, the emulator prediction at any given cosmology retains the large-scale-mode configuration of the shared LHS seed, while the covariance mean represents a cosmic-variance-averaged estimate of the true mean power spectrum.

\begin{figure*}[htbp]
  \centering
  \includegraphics[width=\textwidth]{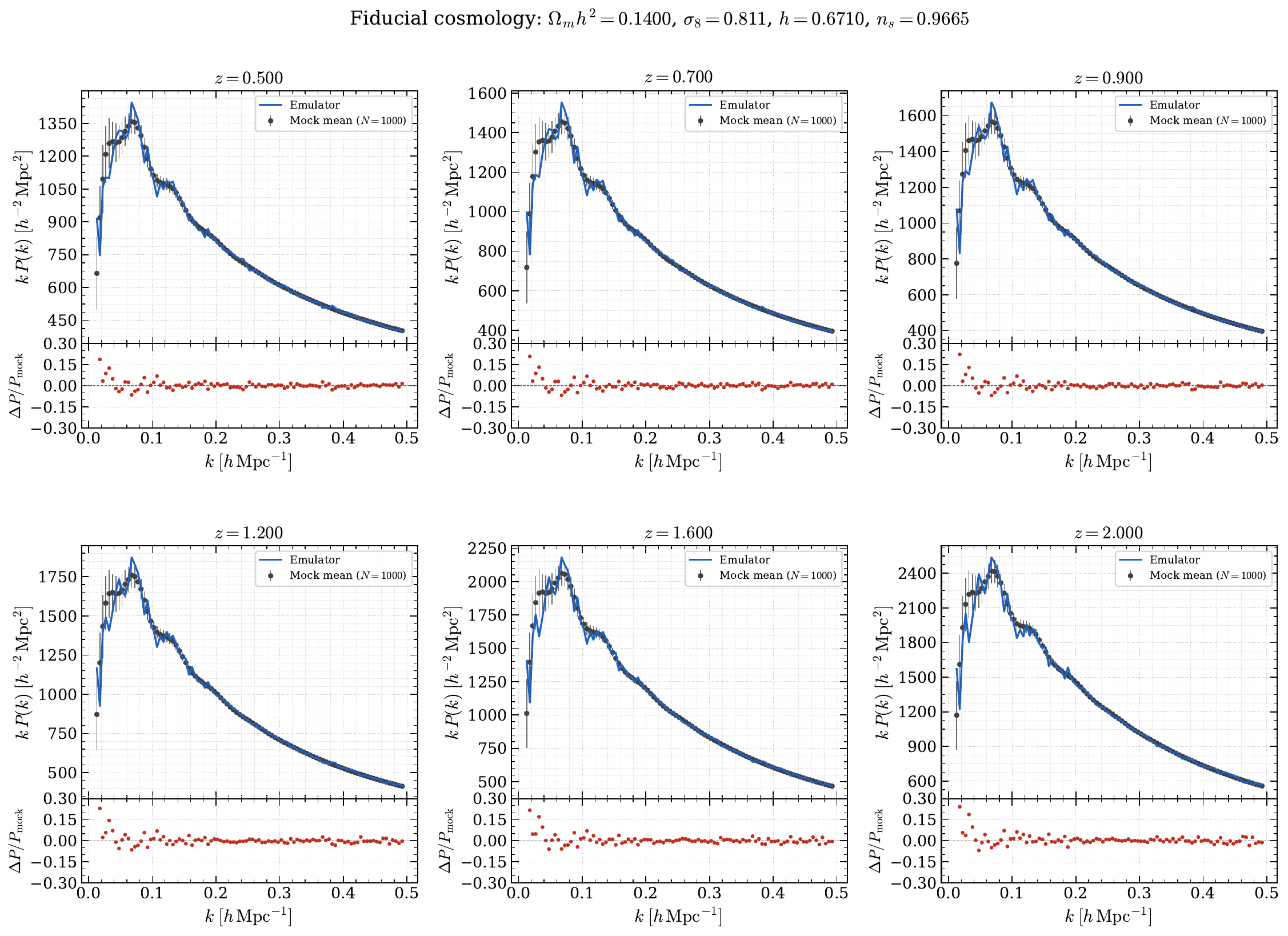}
  \caption{Comparison of the emulator prediction at the fiducial cosmology (blue line) with the mean monopole power spectrum from 1000 covariance-suite realizations (black points with $1\sigma$ error bars) at six redshifts (redshift-space, full halo sample). The lower sub-panels show the fractional difference $\Delta P/P_\mathrm{mock}$. The offsets at $k\lesssim 0.05\,h\,\mathrm{Mpc}^{-1}$ arise because the LHS training suite shares a single random seed, whereas the covariance mean is averaged over 1000 independent seeds (see text). At $k\gtrsim 0.1\,h\,\mathrm{Mpc}^{-1}$ the agreement improves to the few-per-cent level.}
  \label{fig:pred_vs_cov}
\end{figure*}

Figure~\ref{fig:pred_vs_cov} compares the two across six redshifts. The upper sub-panels show the emulator prediction (blue) against the covariance-suite mean (black points with $1\sigma$ error bars). The lower sub-panels display the fractional difference $\Delta P/P_\mathrm{mock}=(P_\mathrm{emu}-P_\mathrm{mock})/P_\mathrm{mock}$. At wavenumbers $k\gtrsim 0.1\,h\,\mathrm{Mpc}^{-1}$, the emulator and the mock mean agree to within a few per cent. At the lowest wavenumbers ($k\lesssim 0.05\,h\,\mathrm{Mpc}^{-1}$), offsets of $10$--$20\%$ are visible and the scatter in the residuals is noticeably larger.

This scale-dependent pattern is a direct and expected consequence of the fixed-seed design of the LHS suite \citep[see, e.g.,][]{Villaescusa_Navarro_2020}. At low $k$, each bin contains only a handful of Fourier modes whose amplitudes and phases are set by the initial random field. Because all 1000 training simulations start from the same seed, the emulator learns the cosmology dependence of the power spectrum conditioned on that particular realization. When compared with the covariance-suite mean, which averages over 1000 independent phase realizations, the residual at low $k$ therefore reflects the difference between one random draw and the population average, namely the cosmic variance of a single $(1024\,h^{-1}\mathrm{Mpc})^3$ box. At higher $k$, the number of modes per bin grows and the single-realization spectrum converges toward the ensemble mean.

The subsequent MCMC analysis shows that these large-scale offsets do not prevent recovery of the fiducial cosmology. The error bars in Fig.~\ref{fig:pred_vs_cov} also show that the observed low-$k$ discrepancies are consistent with the expected uncertainties of the covariance-suite mean, supporting the interpretation that the mismatch is statistical rather than a systematic interpolation failure.

\subsection{MCMC Parameter Constraints}
\label{subsec:mcmc_results}

We now present the MCMC constraints on the six cosmological parameters using the GP emulator as the theory model and the covariance-suite mean as the synthetic data vector. The MCMC is run independently at three representative redshifts, $z=0.5$, $1.0$, and $2.0$, to assess the redshift dependence of the constraining power.

\begin{figure*}[htbp]
  \centering
  \includegraphics[width=0.85\textwidth]{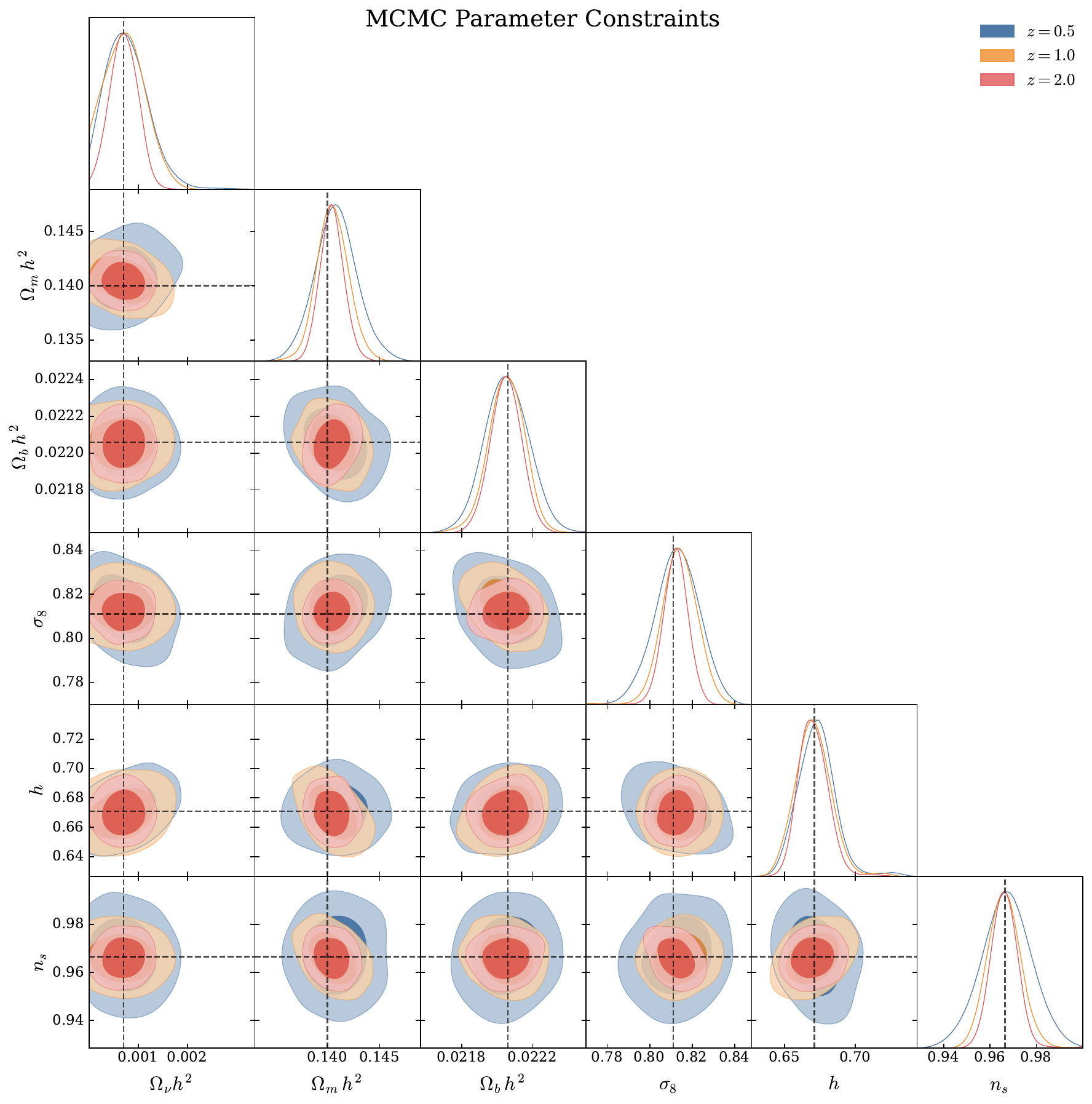}
  \caption{Posterior constraints on the six cosmological parameters from MCMC analyses using the GP emulator as the theory model, with $k_\mathrm{max}=0.50\,h\,\mathrm{Mpc}^{-1}$. Results are shown at three redshifts: $z=0.5$ (blue), $z=1.0$ (orange), and $z=2.0$ (red). The synthetic data vector is the mean monopole from the 1000 covariance-suite realizations (redshift-space, full halo sample). Contours show the 68\% and 95\% credible regions, and the fiducial parameter values (Table~\ref{tab:params}) are marked by cross-hairs.}
  \label{fig:contour_kmax05}
\end{figure*}

Figure~\ref{fig:contour_kmax05} shows the two-dimensional marginalized posterior distributions obtained with $k_\mathrm{max}=0.50\,h\,\mathrm{Mpc}^{-1}$. At all three redshifts the posteriors are unimodal and consistent with the fiducial cosmology (Table~\ref{tab:params}), confirming that the emulator-based likelihood recovers the input parameters without significant bias. The constraints tighten from $z=0.5$ (blue) to $z=2.0$ (red) for several parameters, reflecting the additional leverage provided by higher redshift snapshots where the growth rate sensitivity differs.

The posterior contours in Fig. 7 appear nearly uncorrelated, with little evidence for strong parameter degeneracies. We attribute this in part to the wide range of scales included in the analysis, which provides sufficient information to independently constrain each parameter. Additionally, in the present validation setup, both the synthetic data vector (the mean of 1000 realizations) and the emulator predictions (where the GP noise kernel $\sigma_n^2$ absorbs part of the stochastic scatter) are less noisy than would be the case in a real observational analysis. As a result, the likelihood surface is more sharply peaked than it would be in practice, and degeneracy directions are less prominent. We expect that parameter degeneracies will become more apparent when the emulator is applied to actual survey data, where the data vector is a single noisy realization, observational systematics are present, and additional nuisance parameters (e.g., galaxy bias) are marginalized over.

We note that in the present validation setup, the synthetic data vector is nearly noiseless, while the emulator prediction carries residual stochasticity inherited from individual training simulations. This is the reverse of a standard observational analysis. Nevertheless, the test remains a valid check for systematic biases introduced by the emulation, and the parameter uncertainties are governed by the emulator's effective noise level corresponding to a single simulation volume.

To assess the impact of the mildly nonlinear scales accessed by the emulator, we repeat the analysis with a more conservative scale cut of $k_\mathrm{max}=0.20\,h\,\mathrm{Mpc}^{-1}$, thereby restricting the data vector to the quasi-linear regime. The resulting posteriors are shown in Appendix~\ref{app:mcmc_kmax02}. The contours broaden substantially for all six parameters when the high-$k$ information is excluded. Reducing $k_\mathrm{max}$ from $0.50$ to $0.20\,h\,\mathrm{Mpc}^{-1}$ degrades the marginalized $1\sigma$ constraints by factors of approximately $2$--$4$, depending on the parameter and redshift. This comparison demonstrates the emulator's ability to capture the nonlinear response of the power spectrum to cosmological parameters, thus enabling the extraction of information from scales that would be difficult to model with perturbative approaches alone. The tighter constraints obtained with 
$k_\mathrm{max} = 0.50\;h\,\mathrm{Mpc}^{-1}$ 
confirm that the emulator-based forward model provides a practical method of exploiting the mildly nonlinear regime for neutrino mass inference, which could be of significant use in fully leveraging the statistical power of DESI and other Stage-IV surveys.

\section{Discussion and Conclusions}
\label{sect:conclusion}

We have developed a Gaussian-process emulator for the monopole of the redshift-space halo power spectrum in $\Lambda$CDM cosmologies with massive neutrinos. The emulator is trained on a Latin-hypercube suite of 1000 COLA simulations spanning a six-dimensional cosmological parameter space, with outputs at 11 redshift snapshots over $0.5\leq z\leq 2.0$. From the resulting redshift-space halo catalogues, we measure the shot-noise-subtracted monopole power spectrum over $0.01\leq k\leq 0.50\,h\,\mathrm{Mpc}^{-1}$. In addition, we generate a dedicated covariance suite of 1000 fixed-cosmology realizations, which provides both the covariance matrix for likelihood analysis and a synthetic data vector for end-to-end MCMC tests.

The emulator reaches sub-per-cent to $\sim 2\%$ accuracy on held-out cosmologies across the wavenumbers and redshifts considered (Fig.~\ref{fig:emu_pred}), demonstrating that the per-$k$-bin GP architecture with an ARD kernel captures the scale- and cosmology-dependent response of the monopole, including the characteristic suppression induced by massive neutrinos. When applied as the theory model in an MCMC inference pipeline, the emulator successfully recovers the fiducial cosmology from a synthetic data vector, with constraints that tighten appreciably when the analysis is extended from $k_\mathrm{max}=0.20$ to $0.50\,h\,\mathrm{Mpc}^{-1}$ (Figs.~\ref{fig:contour_kmax05} and \ref{fig:contour_kmax02}). This improvement underscores the value of accurately modelling the mildly nonlinear regime for neutrino-mass constraints.

The main strength of the framework is that it combines physically motivated simulation output with the computational speed required for cosmological parameter inference. By emulating the logarithm of the power spectrum with Gaussian processes, we obtain a fast surrogate model that predicts $P_0(k)$ across cosmological parameter space while naturally returning an estimate of interpolation uncertainty. Standardized inputs, an RBF kernel with automatic relevance determination, and an explicit noise term allow the emulator to capture the scale-dependent sensitivity of the power spectrum to cosmological parameters while smoothing over stochastic fluctuations from sample variance and finite halo number.

At the same time, the present implementation is intentionally focused and therefore has several limitations. First, we consider only the monopole of the redshift-space power spectrum. A fuller exploitation of redshift-space information will require extending the emulator to higher multipoles, especially the quadrupole and hexadecapole. Second, the per-bin GP strategy offers simplicity and computational efficiency, but it does not explicitly model correlations between neighbouring $k$ bins in the emulator itself. Although the smoothness of the physical power spectrum mitigates this issue, it remains an approximation that should be revisited in future high-precision applications. Third, the present shot-noise correction assumes Poisson statistics; this is adequate for the current methodological study, but departures from Poisson behaviour for massive haloes may matter in more detailed analyses.

A further limitation is that the current framework is built at the halo level. Direct application to observed galaxy clustering will require additional ingredients, including a more complete treatment of tracer bias, redshift evolution, and survey-specific observational effects. Likewise, the covariance matrix used here is estimated at a single fiducial cosmology. That simplification is sufficient for the present mock-data demonstration, but future work should assess whether cosmology dependence of the covariance must be incorporated for precision analyses. Additionally, the emulator is trained independently 
at each snapshot redshift and does not interpolate in redshift; 
extending it to treat redshift as a continuous input is left for 
future work.

Despite these caveats, the emulator developed here provides a promising step toward practical neutrino-mass inference from redshift-space clustering. Massive neutrinos imprint a scale- and redshift-dependent suppression of structure growth, and extracting this signal from spectroscopic surveys requires models that move beyond simple linear theory while remaining fast enough for repeated likelihood evaluations. The combination of fast COLA simulations and Gaussian-process emulation offers a useful compromise between physical fidelity and computational efficiency, making this framework well suited to DESI-motivated analyses.

Natural next steps are clear. On the methodological side, the most important extensions are a more detailed parameter-recovery campaign and the inclusion of higher redshift-space multipoles. On the modelling side, incorporating more realistic tracer prescriptions and survey effects will be essential for direct application to observational data. With those developments, the present emulator could become a useful component of future large-scale-structure analyses aimed at precision constraints on neutrino mass and other cosmological parameters.

\begin{acknowledgements}
JG, YF and GBZ are supported by NSFC grant 11925303, and by the CAS Project for Young Scientists in Basic Research (No. YSBR-092). GBZ is also supported by the New Cornerstone Science Foundation through the XPLORER prize.
\end{acknowledgements}

\clearpage
\appendix

\section{Training Diagnostics}
\label{app:training}

In this appendix we present diagnostics of the GP training procedure that complement the validation results shown in Section~\ref{subsec:emu_results}. Both diagnostics are evaluated at the same six redshifts used for the covariance and consistency analyses: $z=0.5$, $0.7$, $0.9$, $1.2$, $1.6$, and $2.0$.

\begin{figure*}[htbp]
  \centering
  \includegraphics[width=\textwidth]{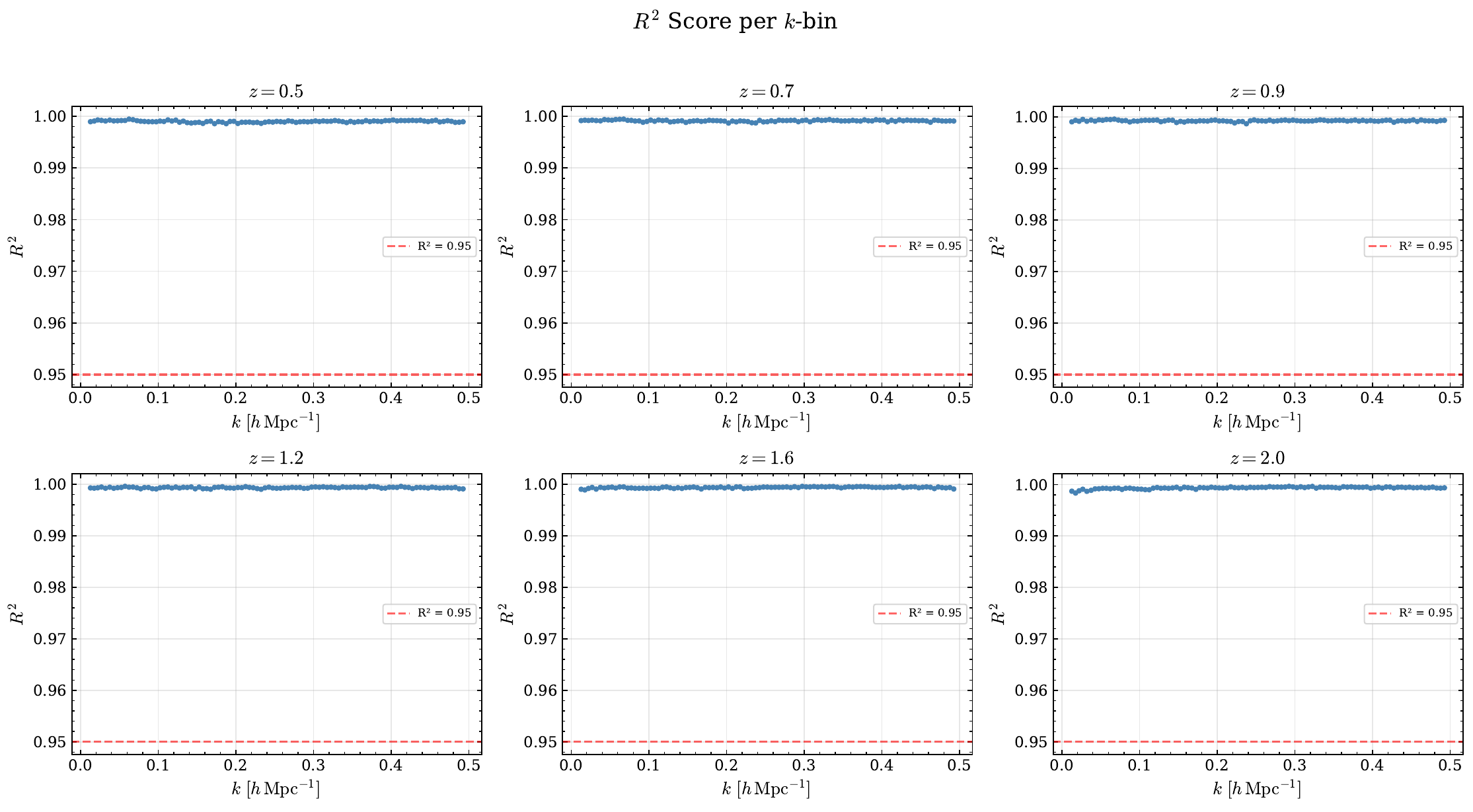}
  \caption{Coefficient of determination $R^2$ evaluated on the validation simulations as a function of wavenumber for the redshift-space monopole emulator (full halo sample) at six redshifts. The dashed red line marks $R^2=0.95$. All $k$ bins at all redshifts exceed $R^2>0.995$, confirming uniformly high emulator accuracy across the full scale and redshift range.}
  \label{fig:r2}
\end{figure*}

Figure~\ref{fig:r2} shows the $R^2$ score on the validation set as a function of wavenumber at each redshift. Across the full range $0.01\leq k\leq 0.50\,h\,\mathrm{Mpc}^{-1}$ and at all six redshifts, every $k$-bin GP achieves $R^2>0.995$, well above the $R^2=0.95$ threshold indicated by the dashed red line. This confirms that the emulator captures the cosmology dependence of the monopole with uniformly high accuracy.

\begin{figure*}[htbp]
  \centering
  \includegraphics[width=\textwidth]{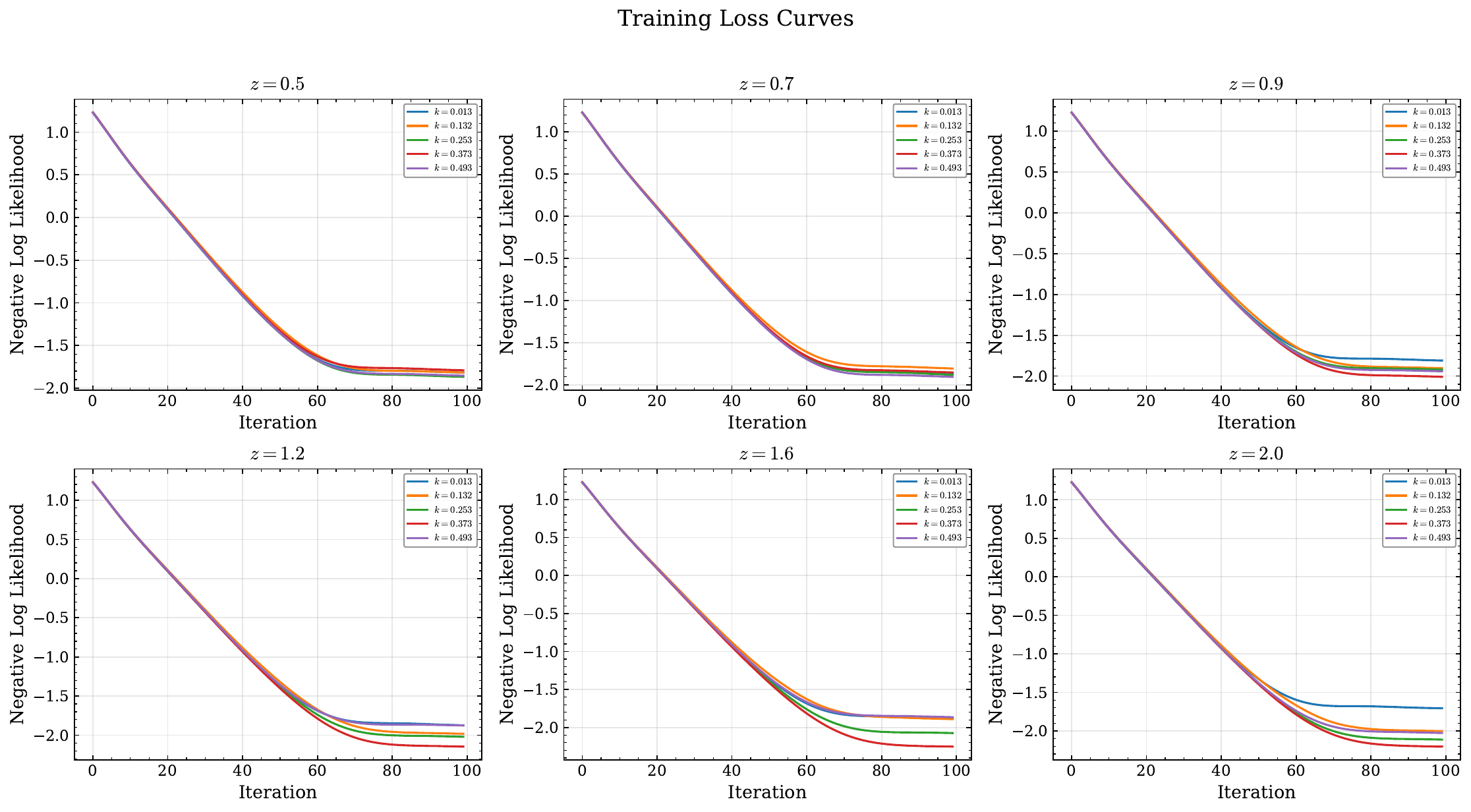}
  \caption{Negative log marginal likelihood as a function of training iteration for five representative $k$ bins ($k=0.013$, $0.132$, $0.253$, $0.373$, $0.493\,h\,\mathrm{Mpc}^{-1}$) at six redshifts. All curves converge within $\sim 80$ iterations.}
  \label{fig:loss}
\end{figure*}

Figure~\ref{fig:loss} displays the negative log marginal likelihood (Eq.~\ref{eq:lml}) as a function of Adam iteration for five representative $k$ bins at each redshift. Convergence is reached within $\sim 80$ iterations for all bins.

\section{MCMC Comparison}
\label{app:mcmc_kmax02}

In this appendix we show the constraints obtained with the more conservative cut $k_\mathrm{max}=0.20\,h\,\mathrm{Mpc}^{-1}$. Relative to the main analysis with $k_\mathrm{max}=0.50\,h\,\mathrm{Mpc}^{-1}$, all contours broaden substantially, and several parameters that are tightly constrained in the fiducial run now occupy a much larger fraction of the prior volume.

\begin{figure*}[htbp]
  \centering
  \includegraphics[width=0.85\textwidth]{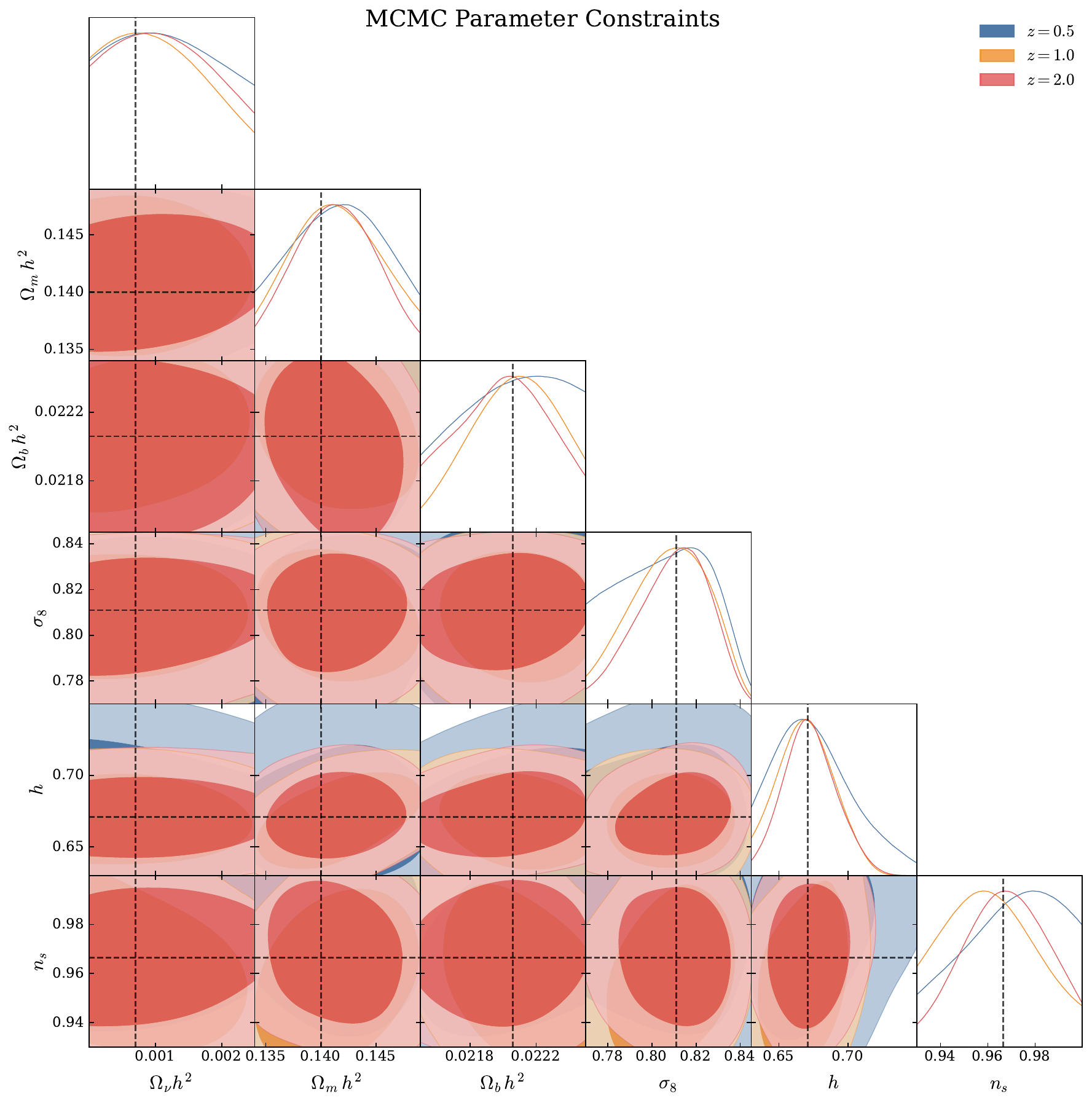}
  \caption{Same as Fig.~\ref{fig:contour_kmax05} but with $k_\mathrm{max} = 0.20\;h\,\mathrm{Mpc}^{-1}$. The broader contours compared with those in Fig.~\ref{fig:contour_kmax05} demonstrate the significant additional constraining power gained by extending the analysis to mildly nonlinear scales, with the marginalized $1\sigma$ constraints degrading by factors of approximately $2$--$4$ across all six parameters.}
  \label{fig:contour_kmax02}
\end{figure*}

\clearpage
\label{lastpage}

\bibliographystyle{raa}
\bibliography{bibtex}
\end{document}